\documentstyle[11pt,paspconf]{article}

\begin{document}

\title{Spatially Resolved Kinematics of Distant Galaxies}

\author{Matthew A. Bershady}
\affil{Astronomy \& Astrophysics Department, Penn State University,
University Park, PA, 16802}



\setlength{\unitlength}{1in}
\begin{picture}(5,0)(0,-3)
\put(-0.8,-0.7){\makebox(5,0){To appear in: Proceedings of {\it Dark and Visible Matter in Galaxies and Cosmological Implications,}}}
\put(-1.1,-0.9){\makebox(5,0){ (Sesto Pusteria, Italy, 2-5 July 1996), eds. M. Persic, P. Salucci (ASP Conference Series)}}
\end{picture}

\def\etal{{\sl et al.}}

\begin{abstract}


A Tully-Fisher (TF) relation from H$\alpha$ rotation curves of 19
luminous, star-forming galaxies reveals there is little evidence for
evolution in the mass-to-light ratio ($M/L$) of these galaxies to
$z\sim0.3$. The near-infrared Tolman surface-brightness test for other
luminous galaxies indicates their luminosity also is little changed to
$z\sim0.3$. In this redshift regime, internal velocity -- luminosity
relations like Tully-Fisher may provide a way to measure
q$_0$. Discrepant results from several intermediate redshift
Tully-Fisher surveys, however, must be understood first. One
possibility is that different surveys sample different galaxy types
and TF relations. Alternatively, $M/L$ of some spiral galaxies may
evolve rapidly with look-back time. Larger surveys are needed to
resolve this issue. A different approach is to determine disk and halo
$M/L$ separately. Such measurements, even at low redshift, would be
sensitive to the star formation histories of disks. We outline plans
for $\lambda/\Delta\lambda\sim10000$ integral-field spectroscopy of
relatively face-on spirals using the 9m Hobby-Eberly
Telescope. Rotation curves and disk stellar velocity dispersions can
provide {\it statistical} information about the $M/L$ of dark halos as
well as luminous disks.

\vskip -0.2in

\end{abstract}



\section{Rotation Curves of blue galaxies to $z\sim 1$}

The era is now underway when faint galaxy surveys include estimates of
internal kinematics. Rotation curves and line-widths are routinely
measured on 3-4m class telescopes to $z\sim 0.4$ (Vogt {\it et al.}
1993, Franx 1993, Bershady 1995b, Bender {\it et al.} 1996, Rix {\it
et al.} 1996, Simard \& Pritchet 1996), and to $z\sim 1$ with 10m
class telescopes (Vogt {\it et al.} 1996a,b). A pressing question is
whether current kinematic surveys show evidence for rapid evolution in
$M/L$, or if internal velocity -- luminosity relations can be used to
construct standard candles for cosmological measurements (Kron 1986,
van der Kruit \& Pickles 1988). How $M/L$ changes with redshift, say
as a function of total mass, will depend in part on the amount and
composition of dark matter.

A number of recent pilot surveys have attempted to construct
Tully-Fisher (TF) relations at intermediate redshifts with a dismaying
range of results as interpreted by departures from local Tully-Fisher
relations. Our survey of H$\alpha$ rotation curves for blue, luminous
galaxies reveals little evidence for evolution in $M/L$ to $z \sim
0.3$ (Bershady 1995b, Bershady, Mihos, \& Koo 1997). The left panel of
figure 1 shows our sample closely matches the $B$-band TF relation of
local calibrators (Pierce \& Tully 1992) assuming H$_0$=75 km s$^{-1}$
Mpc$^{-1}$. A similar agreement is found in $R, I,$ and $H$-bands. A
small offset is observed to increase systematically from $I$
(negligible) to $B$ ($\sim$+0.3 mag). The trend is consistent with
underestimates of disk inclination and hence under-corrections for
internal reddening in our sample. Better inclination estimates are in
progress.

\begin{figure} 
\plotfiddle{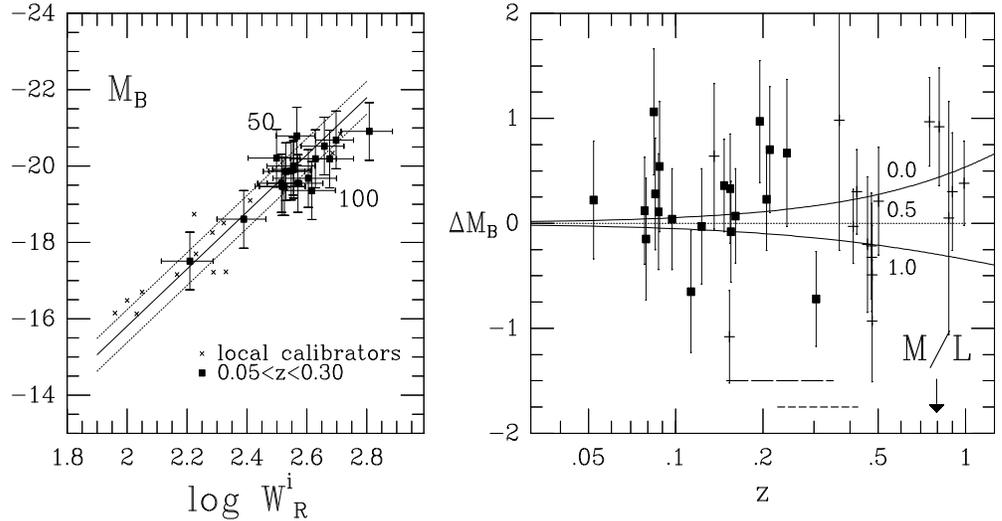}{1.6in}{-90}{56}{60}{-220}{202}
\vskip 0.9in 
\caption{\hsize 5.25in \baselineskip 0.15in \underline{\bf Left Panel:}
$B$ band luminosity vs. line-width for 19 of 21 targets in our sample
with successfully measured H$\alpha$ rotation curves. These are
compared to local calibrators and their regressions from Pierce \&
Tully (1992) assuming $H_0 = 75$ km s$^{-1}$Mpc$^{-1}$. 21-cm
line-widths, corrected for inclination, are estimated from the
rotation curves. Luminosities are corrected for internal reddening
following Tully \& Foque (1985) for consistency with Pierce \& Tully
(1992). ``Error bars'' in luminosity represent different assumptions
for H$_0$: upper (50), lower (100). Photometric errors are $<$3\%.
\underline{\bf Right Panel:} Deviations in $B$ band magnitudes for
current samples at intermediate redshift from various fiducial ($z=0$)
Tully-Fisher relations assuming q$_0$=0.5, H$_0$=75 km
s$^{-1}$Mpc$^{-1}$). Symbols: filled squares, our sample (Bershady,
Mihos \& Koo 1997); plusses, Vogt {\it et al.} (1996a,b); long-dashed
line, average for Rix {\it et al.} (1996); short-dashed line, average
for Simard \& Pritchet (1996). Curves represent secular changes for
different q$_0$ (labeled); arrow represents secular evolution
(decrease) in $M/L$. Note the significant scatter within individual
surveys as well as the very discrepant results between some of the
surveys.}
\vskip -0.15in 
\end{figure}

More relevant to the issue of evolution is an examination of
differential departures from a fiducial TF relation vs. redshift
(right panel of figure 1). Our sample together with those extending to
$z\sim 1$ from the Keck telescope (Vogt {\it et al.} 1996a,b) show
little evidence for dramatic changes in $M/L$. It is clear that with
these small samples, little can be determined about
q$_0$. However, two other samples (Rix {\it et al.} 1996, using [O~II]
$\lambda 3727$ line-widths; and Simard \& Pritchet 1996, using [O~II]
$\lambda 3727$ rotation curves) find evidence for brightening of 1.5-2
mag by z=0.25-0.4. Why are there such differences between surveys
spanning comparable redshifts?

The above surveys consist of samples selected to optimize telescope
efficiency. Our strategy (see Bershady 1995b) was, in short, to select
galaxies that were optically blue (expected to have strong line
emission), but luminous in the near-infrared (expected to be large and
have rapid rotation). The Rix {\it et al.} (1996) and Simard \&
Pritchet (1996) surveys relax the luminosity criteria but have
(different) color or emission line-strength criteria. The samples of
Vogt {\it et al.} (1996a,b) are selected on the basis of visual
morphology from HST WFPC-2 images. For a variety of reasons, either
the characteristic luminosity, rest-frame color, or redshifts of
these four surveys are different (figure 2). Rix {\it et al.}'s (1996)
sample is $\sim$1 mag lower in luminosity than the others; Simard \&
Pritchet's (1996) sample is intermediate in redshift between ours and
Vogt {\it et al.}'s (1996a,b); our sample is redder (in the
rest-frame) than the other three.

\begin{figure}
\plotfiddle{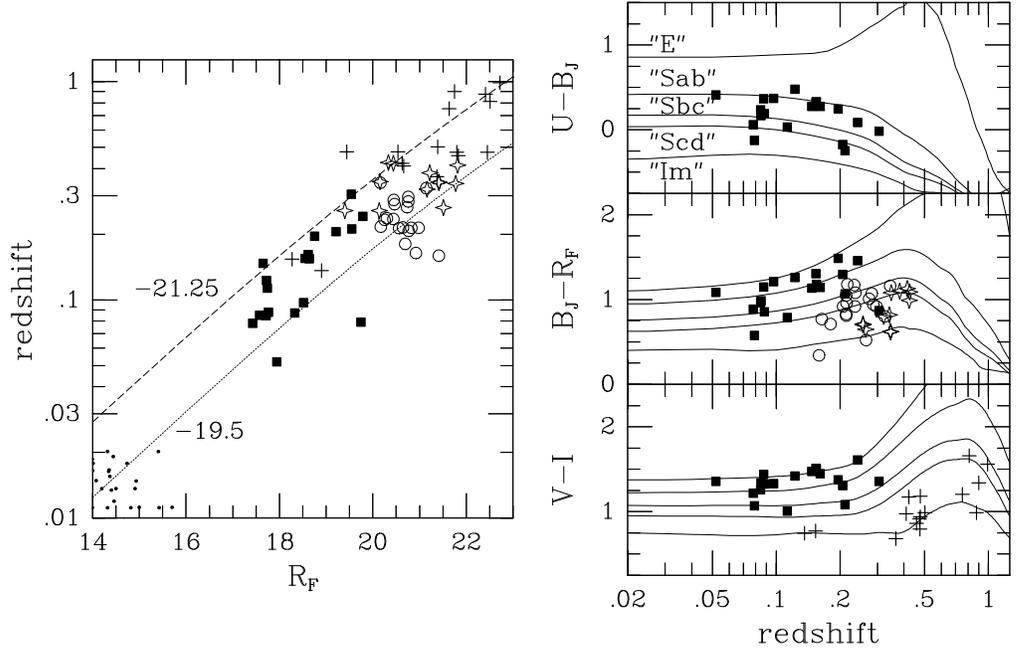}{0in}{-90}{60}{60}{-215}{65}
\vskip 3.15in
\caption{\hsize 5.25in \baselineskip 0.15in Scope and differences
between current intermediate redshift TF surveys. \underline{\bf Left
panel:} Hubble diagram in the $R_F$-band. Diagonal lines correspond to
constant luminosity (q$_0$=0.5, H$_0$=75 km s$^{-1}$Mpc$^{-1}$, and
Sbc $k$-corrections), where M$_{R_F}$=-21.25 is $\sim$ M$^*$. Symbols:
filled squares are 19 galaxies from Bershady, Mihos \& Koo (1997);
open circles are 19 of 24 galaxies with good [O~II] line-widths from
Rix \etal\ (1996); plusses are from Vogt \etal\ (1996a,b; $R_F$
magnitudes are estimated from $V$ and $I$); open diamonds are 12 of 24
``kinematically normal'' galaxies from Simard \& Pritchet (1996; $R_F$
magnitudes are estimated from $g$ and $r$). For comparison, the most
distant portion of Mathewson \etal's (1992) local sample is shown
(dots). \underline{\bf Right panel:} Color vs. redshift. Colors
characteristic of different Hubble types are plotted for
reference. Note (1) the different range of luminosity, rest-frame
color, or redshift for each survey; and (2) the small number and
substantially incomplete coverage in luminosity, type, and redshift
for all surveys.}
\vskip -0.15in
\end{figure}

Selection based on luminosity, color, line-strength or morphology is
not a problem for differential measurements with redshift as long as
the selection is well-defined over all redshifts. What is difficult,
however, is comparing samples selected with different criteria,
particularly because internal velocity -- luminosity relations may
systematically vary with galaxy type (e.g. Rubin {\it et al.}
1985). Alternatively, galaxies may evolve differently depending on
mass or other physical attributes. It is critical to distinguish
between the above two possibilities. Most current samples, however,
are either too small or probe too narrow a range of redshift to {\it
internally} measure changes with redshift.

\begin{figure}
\plotfiddle{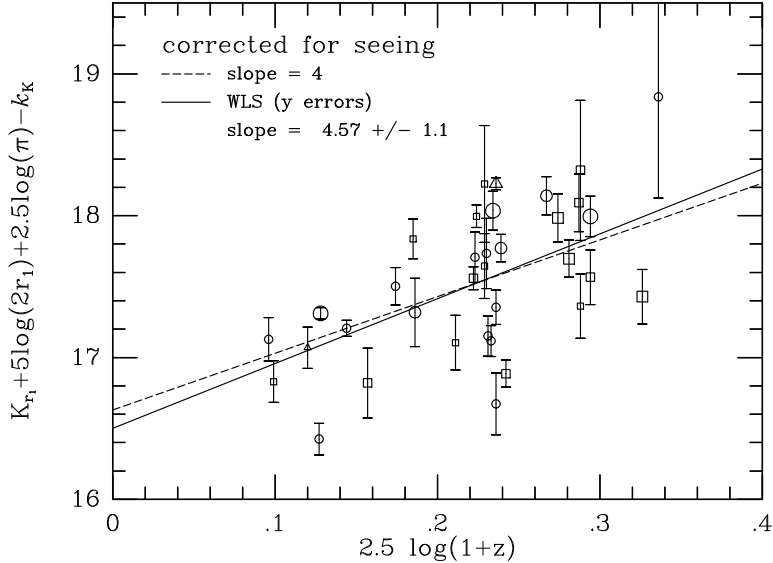}{1.75in}{0}{60}{60}{-180}{-298}
\vskip 1.0in
\caption{\hsize 5.25in \baselineskip 0.15in The Tolman
surface-brightness test in the $K$-band for field galaxies with
$M_K<-24.75$ (H$_0$=km s$^{-1}$Mpc$^{-1}$) from Bershady {\it et al.}
(1994). The luminosity limit is brighter than that for our rotation
curve sample in figure 1, and there is no color selection.  Symbols
refer to spectral types from Bershady (1995a): {\it bm} (triangles),
{\it am} or {\it fm} (squares), {\it gm} (circles). Surface-brightness
is measured as an average within twice the first moment of the light
profile ($r_1$), and is $k$-corrected (Bershady 1995a). Size
measurements have been corrected on the basis of a grid of synthetic
galaxy images for a range of disk-to-bulge ratios, inclinations and
sizes, which are aberrated for seeing. The aberrated model best
matching an observed galaxy's size, inclination, and image
concentration is used to estimate corrections. The corrections are
small. The dashed line (slope of 4) indicates expectations for no
luminosity evolution, assuming the expansion is real. The solid line
is the best fit using a weighted linear least-squares (WLS) regression
from Akritas \& Bershady (1996), which accounts for the intrinsic
scatter in surface-brightness as a function of galaxy type and
luminosity. The observed slope is consistent with no evolution,
although the uncertainty is large given the small sample size.}
\vskip -0.15in
\end{figure}

More troublesome are potential biases with redshift in the kinematic
measurements themselves. For example, interpretation of spatially
integrated line-widths suffers from the ambiguity of not knowing the
spatial distribution of the line-emitting material. In the optical, a
centrally concentrated star-burst compared to a blue, star-forming
disk may have drastically different line-widths even though their
gravitational potentials are similar. Rix {\it et al.}  (1996) have
presented the most comprehensive attempt to treat this problem to
date. To complicate matters, it is likely that there are systematic
differences between the spatial distribution of [O~II] $\lambda 3727$
and H$\alpha$ in galaxies. A comprehensive comparison of H$\alpha$ and
[O~II] $\lambda 3727$ even within local galaxies is lacking.  More
subtle are the effects of spatial resolution on spatially resolved
rotation curve measurements. For example, how does the apparent
maximum rotation velocity change for identical galaxies observed at
different distances but in the same seeing? One expects such effects
are small if the galaxies remain larger than the seeing disk; still,
this question must be answered in detail. Careful examination of all
of the above questions is critical because of the steep slope in the
relation between line-width and luminosity.

The future success of using the TF relation to study galaxy evolution
and cosmology depends on two additional factors. First, samples must
be assembled intelligently to make internally-consistent comparisons
over substantial ranges of redshift.  Surveys must be large enough to
construct template relations at different redshifts and account
empirically for internal absorption, as is now becoming possible
locally (e.g. Giovanelli {\it et al.} 1995). Second, independent
assessments of luminosity evolution must be available because
deviations from the power-law TF relation are degenerate for changes
in $M$ and $L$. One possibility is the Tolman surface-brightness test,
an example of which is illustrated in figure 3. While assumptions must
be made about evolution in size, the beauty of this test is that
surface-brightness, like line-width, is curvature independent. For the
most luminous galaxies from Bershady {\it et al.} (1994), figure~3
indicates there is little evidence for brightening in the $K$-band
surface-brightness to $z \sim 0.3$. One reasonable interpretation is
that the luminous mass has changed little in luminous galaxies over
this look-back time. Note that the sample in figure 3 is not the same
as our (rotation curve) sample in figure 1, but illustrates the
concept. Such studies {\it in the field} are easily extendible --
using 10m-class telescopes or the Hubble Space Telescope -- to include
TF samples at $z\sim 1$ or lower luminosities. The near-infrared is
advantageous because $k$-corrections are well-defined to high
redshifts, vary little between galaxy types, and presumably trace
stellar mass.

\section{Disk Kinematics at low redshifts}

Global kinematic measurements, such as rotation curves, are a rather
blunt tool for probing dark {\it and} luminous matter in galaxies
because they are sensitive to the total mass. For spirals, the $M/L$
of the disk is likely to vary much more rapidly with look-back time
than the halo $M/L$. Yet the disk constitutes only a small portion of
the total mass, while providing substantially to the total light.
Hence it would be desirable to independently measure the masses of the
disk and halo. This is possible for spirals if rotation curve and disk
scale length measurements are combined with measurements of the
$z$-component of the disk velocity dispersion ($\sigma_z$) and scale
height (Bahcall \& Casertano 1984). Direct determinations of disk and
total masses would alleviate the assumptions in most disk/halo
decompositions, namely maximal disks and constant disk $M/L$ (e.g. van
Albada {\it et al.} 1985, Sackett 1995).

\begin{figure}
\plotfiddle{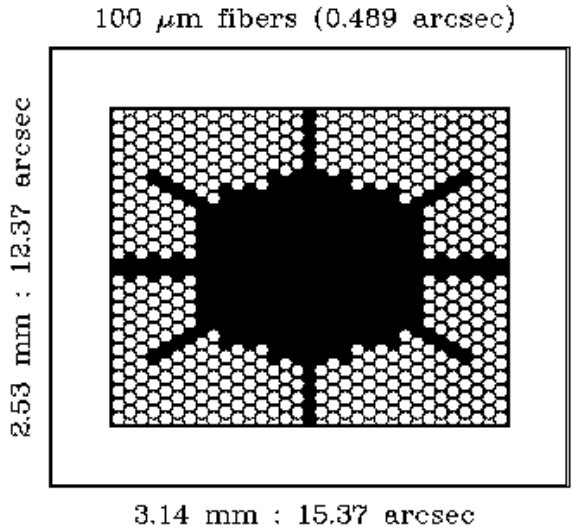}{2.0in}{-90}{65}{65}{-250}{250}
\vskip 0.5in
\caption{\hsize 5.25in \baselineskip 0.15in One of two integral field
fiber array designs for galaxy kinematic studies with the HET's
Medium Resolution Spectrograph (MRS). The 15 arcsec field is filled
with $\sim$0.5 arcsec diameter fibers. Dark fibers go through to the
spectrograph; remaining open fibers are for packing. A second array
uses 1 arcsec fibers to span a 30 arcsecs.  The filling ratio is 75\%
in the coherent regions. Ten sky fibers are attached to the support
structure (not shown). Note that diagonal axes are at
$\pm$30$^{\circ}$ from the major axis.}
\vskip -0.15in
\end{figure}

\begin{figure}
\plotfiddle{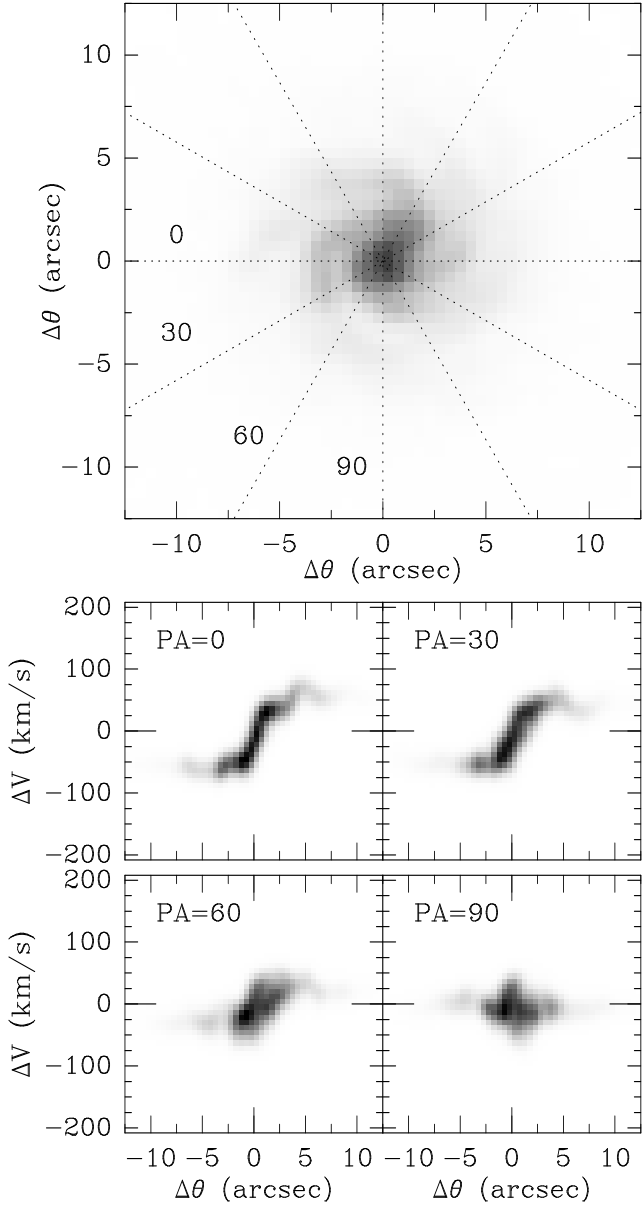}{6.5in}{0}{70}{70}{-200}{20}
\vskip -0.6in
\caption{\hsize 5.25in \baselineskip 0.15in Simulation of a spiral
galaxy observed at $z=0.2$ in 1 arcsec seeing (FWHM) with a fiber
array with 0.5 arcsec fibers and spectral resolution of 10,000
(Bershady \& Mihos 1997). The galaxy has 250 km/s maximum rotation
velocity, 3.75 h$^{-1}$ kpc disk scale length, and inclination of
15$^{\circ}$. Top panel: two-dimensional spatial image is
reconstructed in H$\alpha$. Four bottom panels: spectra centered on
H$\alpha$ are extracted at position angles of 0, 30, 60, and 90
degrees from the galaxy major axis. Noise has not been added. Note
that (1) even at 15 degrees inclination, rotation is easily detected;
(2) a misalignment of the major axis would be easily detected by the
presence of rotation in the minor axis; (3) while the simulation is
for line-emission, schematically it shows how stellar absorption
line-widths can be measured over a large number of resolution elements.}
\vskip -0.15in
\end{figure}

How would such measurement be made? An important issue to consider is
disk inclination. In order to maximize projection, face-on galaxies
are preferable for measuring $\sigma_z$ (which is small), while more
inclined systems are preferable for rotation in order to minimize
corrections. Disk scale height measurements must be made for highly
inclined systems. One option is to survey similar galaxies with a
range of disk inclinations. To separate $\sigma_z$ from the other
components of the disk velocity dispersion ellipsoid in inclined disks
requires observations at multiple position angles. Such
observations will also provide sensitivity to (i) non-circular motions
(Franx \& de Zeeuw 1992); (ii) the radial dispersion $\sigma_R$, which
can be used to explore disk stability (Toomre 1964); and (iii)
kinematic estimates of disk inclination (e.g. van der Kruit \&
Allen 1978).


Can such measurements be made? Disks generally have low surface
brightness and velocity dispersions of few $\times$ 10 km/s, requiring
spectral resolutions of $\sim$8,000-10,000. There have been a number
of measurements of disk velocity dispersions in nearby galaxies
(e.g. recently Bottema 1993, and references therein), but they have
proven difficult on 4m-class telescopes. However, integral field
spectroscopy on 10m-class telescopes would make such measurements
quite feasible. In addition to the increased light gathering power of
a 10m telescope, multiple position angles could be observed
simultaneously. We are designing integral field units for the 9m
Hobby-Eberly Telescope's (HET) Medium Resolution Spectrograph
specifically for observing disk kinematics. One example of an integral
fields unit for the HET is illustrated in figure 4. In two hours the
HET can measure a disk velocity dispersion at $z$=0.2 for what would
take 6 hours at $z$=0 on a 4m telescope. A simulation of a large
spiral galaxy observed with an integral field unit at $z$=0.2 is shown
in figure 5.  A second unit is designed for galaxies at redshifts as
low as 0.05.

Even without disk scale height information for individual galaxies,
integral field spectroscopy will enable us to determine the ratio of
rotation speed to disk velocity dispersion in a single observation per
galaxy. To first order, this ratio scales as the ratio of the total
mass to the mass of the disk. Hence it should be possible to
substantially improve our estimates of galaxy disk and halo
masses. This in turn will allow us to better interpret observed
changes in global $M/L$ at intermediate redshifts and the nature of
dark matter.

\acknowledgments I thank C. Mihos and D. Koo for allowing me to
present results of our rotation curve survey and simulations prior to
publication. This research was supported in part by NASA through grant
HF-1028.02-92A from STScI (operated by AURA, Inc. under contract
NAS5-26555).


\end{document}